\documentclass[fleqn,12pt,twoside]{article}
\usepackage{espcrc1}

\usepackage{graphicx}
\usepackage[figuresright]{rotating}

\newcommand{\AmS}{{\protect\the\textfont2
  A\kern-.1667em\lower.5ex\hbox{M}\kern-.125emS}}

\title{ Viscous Corrections to Spectra, Elliptic Flow, and HBT Radii}

\author{Derek Teaney \address[BNL]{Department of Physics, Bldg. 510A, 
        Upton, NY 11973-5000}}
       
\begin{document}

\maketitle

\begin{abstract}
I compute the first viscous correction to the thermal distribution 
function. With this correction,  I calculate the effect of viscosity
on spectra, elliptic flow, and HBT radii.  Indicating the 
breakdown of hydrodynamics, viscous corrections 
become of order one for $p_{T} \sim 1.5\,\mbox{GeV}$.  
Viscous corrections to HBT radii are particularly large
and reduce the outward and longitudinal radii. This reduction is
a direct consequence of the reduction in longitudinal pressure.
\end{abstract}

\section{Viscous Corrections}
Ideal hydrodynamics describes a wide variety of data from heavy 
ion collisions \cite{Teaney1,Kolb1}. In particular ideal
hydrodynamics successfully predicted the observed elliptic flow 
and its dependence on mass, centrality, beam energy, and 
transverse momentum.  Nevertheless, the hydrodynamic approach 
failed in several respects. First, above a transverse momentum 
$p_{T} \sim 1.5\,\mbox{GeV}$ the particle spectra 
deviate from hydrodynamics and approach a 
power law.  Second, HBT radii 
are significantly too large compared to ideal hydrodynamics \cite{BassDumitru}.
Considering the partial success of ideal hydrodynamics, viscous 
corrections may provide a natural explanation for these failures.

For an ideal Bjorken expansion, the entropy per unit space-time rapidity 
$(\tau s)$ is conserved. For a viscous Bjorken  expansion
the entropy per unit rapidity increases as a function of 
proper time \cite{MG84}  
\begin{eqnarray}
   \frac{d ( \tau s) }{d\tau} = 
   \frac{\frac{4}{3} \eta}{\tau T}\;,
\end{eqnarray}
where $\eta$ is the shear viscosity. In this
equation and below we have neglected the bulk viscosity.
For hydrodynamics to be valid, the entropy produced over the
time scale of the expansion (to wit, 
$\tau \frac{\frac{4}{3} \eta}{\tau T}$)  must be small compared to the 
the total entropy ($\tau s$). This leads to the requirement that
\begin{eqnarray}
     \frac{\Gamma _{s}}{\tau} \ll 1\;, 
\end{eqnarray}
where we have defined the {\it sound attenuation length} as
$\Gamma_{s} \equiv \frac{ \frac{4}{3} \eta } {s T}$.  Perturbative
estimates of the shear viscosity in the plasma 
give $\frac{\Gamma_{s}}{\tau} \sim 1 $. 
Below we take $\frac{\Gamma_{s}}{\tau} = \frac{1}{3}$, assuming that
non-perturbative effects shorten equilibration times.

Viscosity modifies the thermal distribution function. This modification
influences the observed particle spectrum and HBT correlations. The formal procedure for determining the viscous correction 
to the thermal distribution function is described in references
\cite{deGroot,Yaffe}. However, the basic form of the viscous correction
can be intuited without calculation. First write 
$f(p) = f_{o}(1 + g(p))$, where $f_{o}(\frac{p\cdot u}{T})$ is the equilibrium
thermal distribution function and $g(p)$ is the 
first viscous correction.   
$g(p)$ is linearly proportional to the spatial gradients in the system. 
Spatial gradients which have no time derivatives in the rest frame  and
are therefore formed with the differential operator 
$\nabla_{\mu} = (g_{\mu\nu} - u_{\mu}u_{\nu})\partial^{\nu}$\,.
For a baryon free fluid, these gradients are $\nabla_{\alpha}T$,  
$\nabla_{\alpha}u^{\alpha}$, and 
$\left\langle \nabla_{\alpha}u_{\beta} \right\rangle$, where  
$\left\langle \nabla_{\alpha}u_{\beta} \right\rangle \equiv 
\nabla_{\alpha}u_{\beta} + \nabla_{\beta}u_{\alpha} - 
\frac{2}{3} \Delta_{\alpha\beta}\nabla_{\gamma}u^{\gamma}$.
$\nabla_{\alpha}T$ can be eliminated in favor of the other 
two spatial gradients using the condition that 
$T^{\mu \nu}u_{\nu} = \epsilon u^{\mu}$ and the ideal equations
of motion. $\nabla_{\alpha}u^{\alpha}$ 
leads ultimately to a bulk viscosity and will be neglected in 
what follows. Finally, $\left\langle \nabla_{\alpha}u_{\beta} \right\rangle$ 
leads to a shear viscosity. If $g(p)$ is
restricted to be a polynomial of degree less than two, then the functional
form of the viscous correction  is completely determined,
\begin{eqnarray}
\label{correction}
   f = f_{o}(1 + \frac{C}{T^3} p^{\alpha}p^{\beta} 
   \left\langle \nabla_{\alpha}u_{\beta} \right\rangle)\;.
\end{eqnarray}   
For a Boltzmann gas this is the form of the viscous correction adopted
in this work. For Bose and  Fermi
gasses the ideal distribution function in Eq. \ref{correction}  
is replaced with $f_{o}(1 \pm f_{o})$ \cite{Yaffe}.

The coefficient $C$ is directly related to the sound attenuation
length. Indeed, using the distribution function in Eq. \ref{correction} 
to determine the stress energy
tensor, yields a relationship between the shear viscosity $\eta$ and the coefficient $C$,
\begin{eqnarray}
\label{tensor}
   T^{\mu\nu}= T^{\mu\nu}_{o} + \eta 
   \left\langle \nabla^{\mu}u^{\nu} \right\rangle  = 
             \int d^3p \, \frac{p^{\mu} p^{\nu}}{E} \,
             f_{o}(1 + \frac{C}{T^3} p^{\alpha}p^{\beta} 
   \left\langle \nabla_{\alpha}u_{\beta} \right\rangle) \; .
\end{eqnarray}
For a Boltzmann gas, Eq. \ref{tensor} yields $C=\frac{\eta}{s}$. 

The thermal distribution function is now completely determined. In 
the next section this correction is used to calculate corrections
to the observables used in heavy ion collisions.

\section{Corrections to Spectra, Elliptic Flow, HBT Radii}

To quantify the effect of viscous corrections on spectra and  HBT
radii, I generalize  the blast wave model.
In the blast wave model used here,  the matter undergoes a boost 
invariant Bjorken expansion
and decouples at a proper time of $\tau_{o}=6.5\:\mbox{Fm}$ at a temperature 
of $T_{o}=160\:\mbox{MeV}$. 
The matter is distributed  
uniformly up to a radius of $R_{o}=10.0\:\mbox{Fm}$, with  velocity profile up to a maximum velocity $u^{r}_{o}=0.5\,c$
\footnote{In contrast to common practice $u^{r} = \gamma v^{r}$ is 
linearly rising: $u^{r} = u^{r}_{o} \frac{r}{R_{o}}$}.  
The 
blast wave model with these parameters closely models the 
output of a full hydrodynamic simulation \cite{Teaney1,Kolb1} 
and gives a reasonable fit of the data.

\begin{figure}[tb]
\begin{minipage}[t]{75mm}
\hspace{-5mm}
\includegraphics[height=85mm, width=85mm]{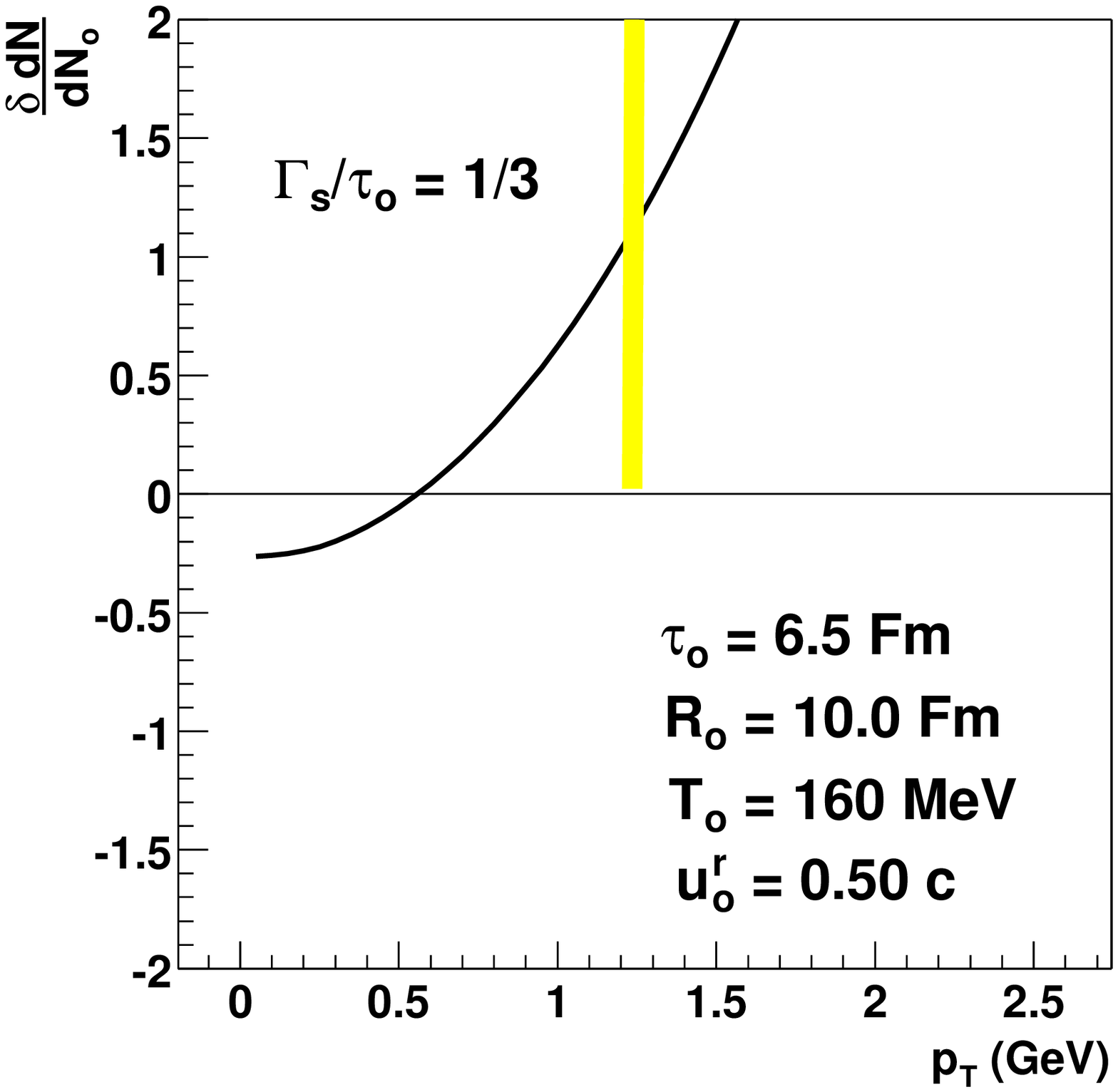}
\vspace{-17mm}
\caption{Viscousity corrected spectrum relative to uncorrected spectrum. 
The band indicates where hydrodynamics breaks down.}
\label{figSpectra}
\vspace{-5mm}
\end{minipage}
\hspace{\fill}
\begin{minipage}[t]{75mm}
\hspace{-5mm}
\includegraphics[height=85mm, width=85mm]{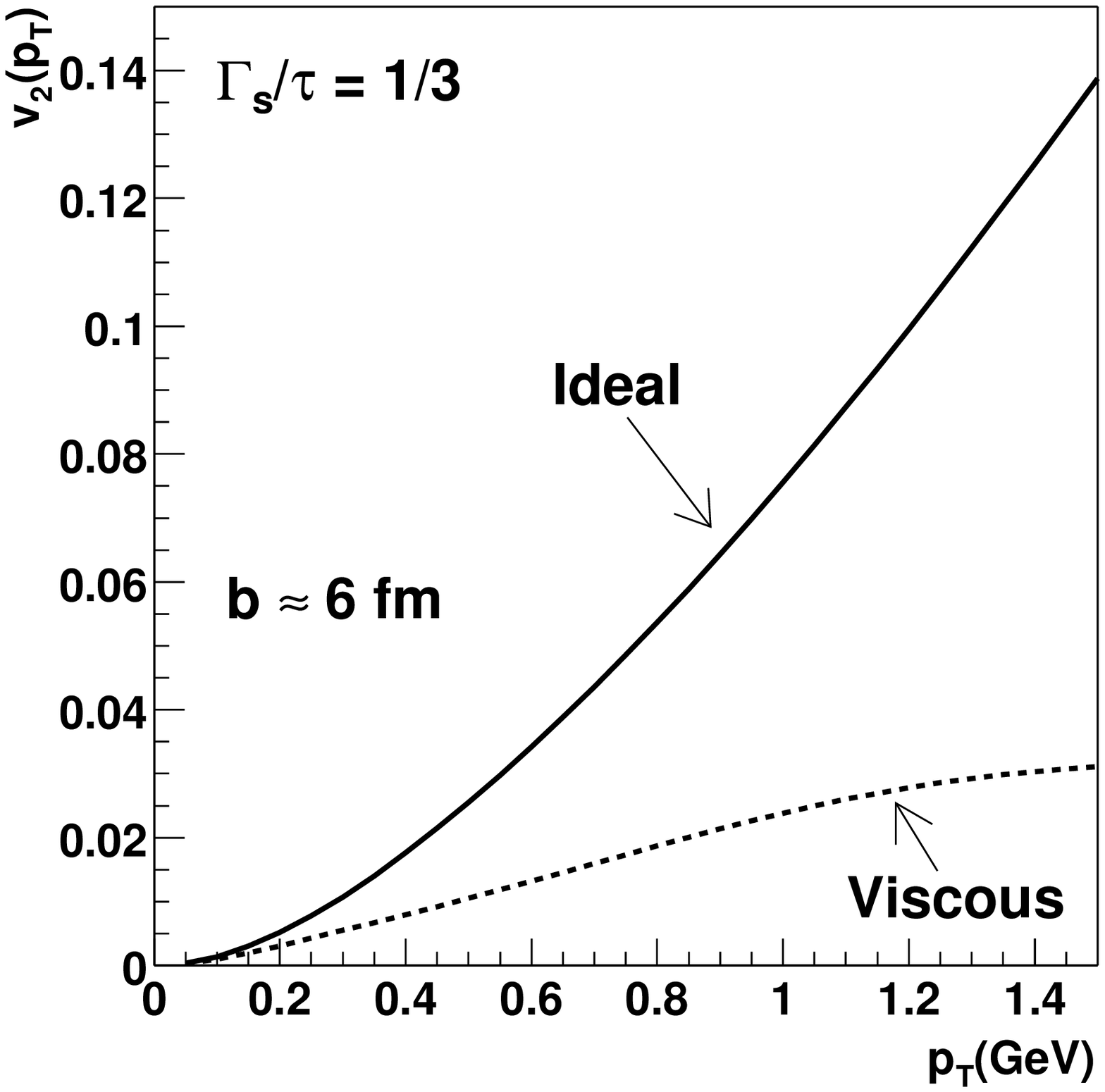}
\vspace{-17mm}
\caption{Elliptic flow as a function transverse  momentum. The
blast wave parameters (see text) are chosen to approximate
a AuAu collision at $b=6\;\mbox{Fm}$.}
\label{figV2}
\end{minipage}
\vspace{-5mm}
\end{figure}

The spectrum of produced
particles is given by the Cooper-Frye formula,
\begin{eqnarray}
\label{EqSpectra}
    \frac{d^{2}N}{d^{2}p_{T}\,dy} &=&  \int p^{\mu} d\Sigma_{\mu}\, f   \\
    dN_{o} + \delta\,dN &=&  \int p^{\mu} d\Sigma_{\mu}\, f_{o}  + \delta f \;. 
\end{eqnarray}
Fig. \ref{figSpectra} shows
the ratio of the correction compared to the ideal spectrum,
$\frac{\delta\,dN}{ dN_{o} }$. 

To understand this figure qualitatively, consider a 
Bjorken expansion of infinitely large nuclei. 
The longitudinal pressure is reduced \cite{MG84}, 
$p_{L}=p-\frac{4}{3}\frac{\eta}{\tau}$.
Because the shear tensor is traceless, 
the {\it transverse} pressure is {\it increased}, 
$p_{T} = p + \frac{2}{3}\frac{\eta}{\tau}$. Thus, the matter 
distribution is pushed out to larger $p_{T}$ by the shear in 
the longitudinal direction. More mathematically, 
the ratio of the corrected spectrum to the uncorrected spectrum is 
given by,
\begin{eqnarray}
\label{spectra}
  \frac{\delta\,dN}{ dN_{o} }  &=& \frac{\Gamma_s} {4\tau}  
  \left\{    \left( \frac{p_T}{T}
             \right)^2  - 
             \left( \frac{m_T}{T} 
             \right)^2 
             \frac{1}{2} 
             \left( \frac{
                           K_3(\frac{m_T}{T})
                         }{ 
                           K_1(\frac{m_T}{T}) 
                         } -1
             \right) 
  \right\}\;.
\end{eqnarray}
For large $p_{T}$ we find,  
$\frac{\delta\,dN}{ dN_{o} } \approx \frac{\Gamma_s}{4\tau}  
 \left( \frac{p_T}{T} \right)^2 $. Eq. \ref{spectra} reproduces
the shape and dependence of the full viscous blast wave calculation 
shown in Fig. \ref{figSpectra}.   

Viscous  corrections become of order one when the $p_{T}$ of the particle
approaches $1.4$ GeV.  This signals the breakdown of the hydrodynamic 
approach. In fact, ideal hydrodynamics generally fails to reproduce
the single particle  spectra above $p_{T}$ of 1.5 GeV. Viscosity
provides a ready explanation for this breakdown.

In non-central collisions elliptic flow is calculated using
the spectrum indicated in Eq. \ref{EqSpectra}. In non-central collisions,
the matter is assumed to have a cylindrical distribution 
but the flow velocity has an elliptic component,
$u^{r}(r,\phi) = u^{r}_{o}\frac{r}{R_{o}}(1 + 2 u_{2} \cos(\phi))$.  
For non-central collisions the parameters are: $u_{o}=0.5, u_{2}=0.1, 
R_{o}=6\;\mbox{Fm}$ and $\tau_{o}=4.0\;\mbox{Fm}$. As illustrated
in Fig. \ref{figV2}, viscosity reduces elliptic flow by a factor of three. 

\begin{figure}[tb]
\begin{minipage}[t]{75mm}
\hspace{-5mm}
\includegraphics[height=85mm, width=85mm]{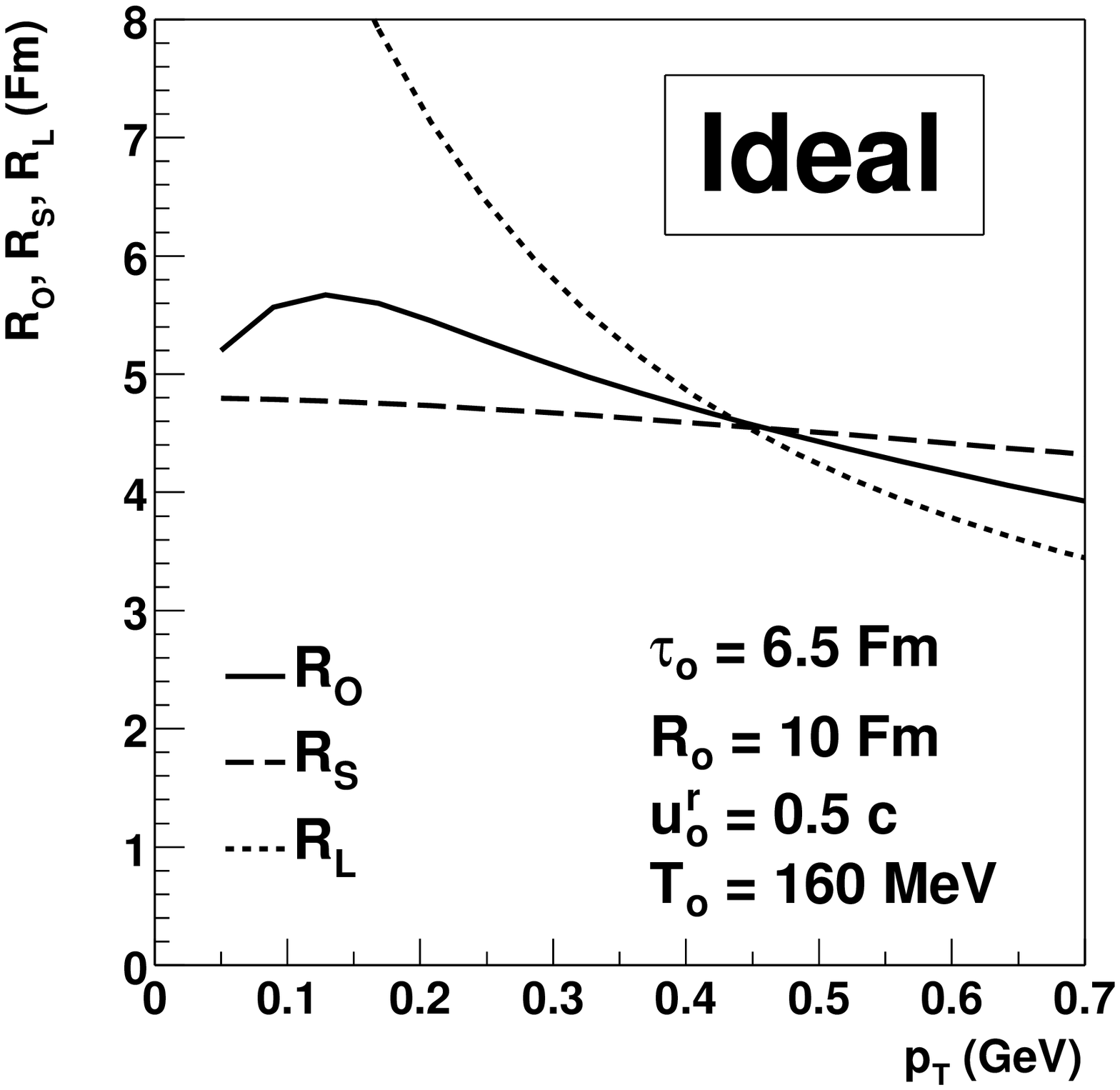}
\vspace{-17mm}
\caption{HBT radii for the ideal blast wave model described
in the text.}
\label{IdealHBT}
\vspace{-5mm}
\end{minipage}
\hspace{\fill}
\begin{minipage}[t]{75mm}
\hspace{-5mm}
\includegraphics[height=85mm, width=85mm]{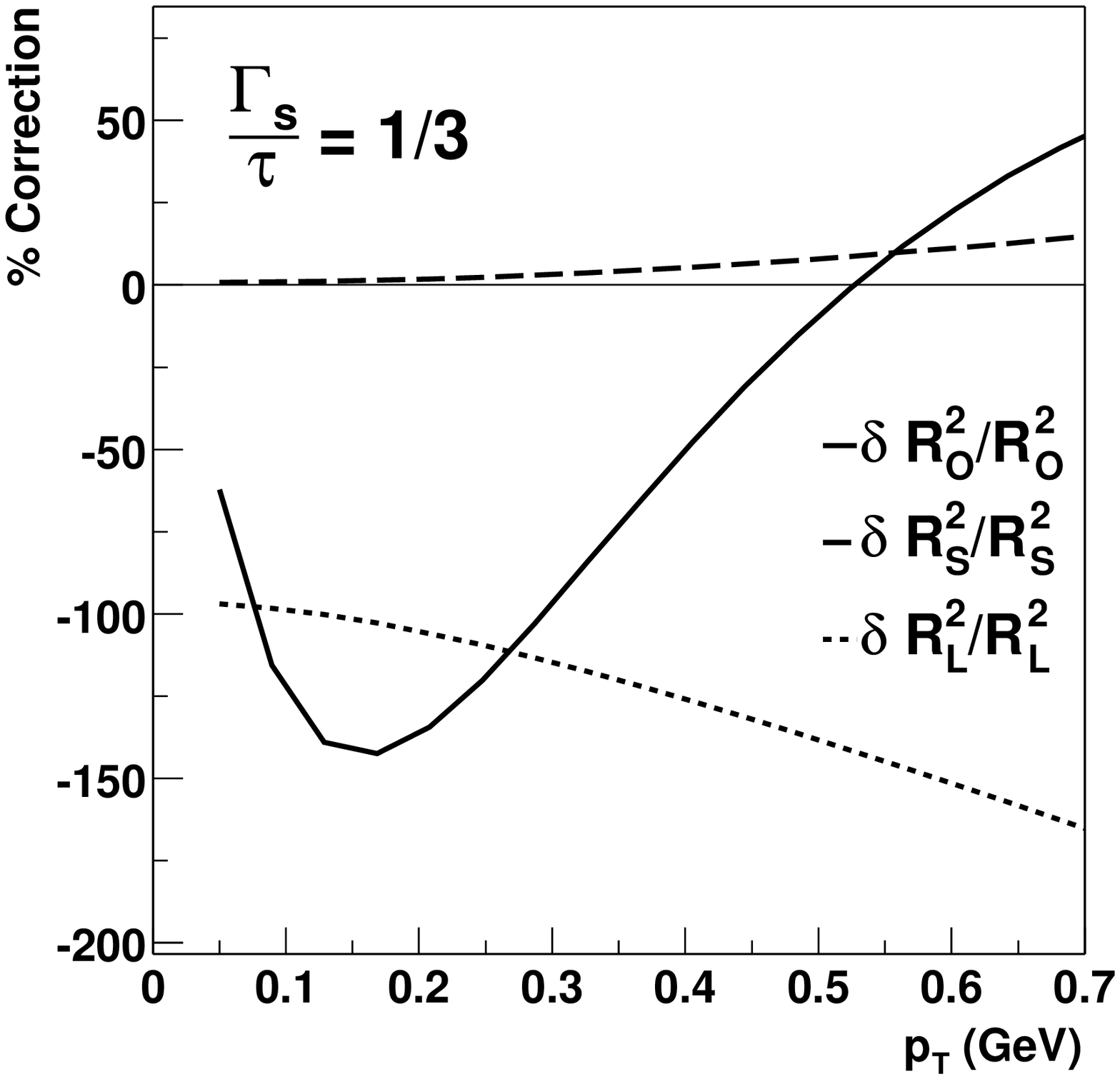}
\vspace{-17mm}
\caption{Viscous corrections to the ideal HBT radii illustrated in Fig. \ref{IdealHBT}.}
\label{VisHBTCorrections}
\end{minipage}
\vspace{-5mm}
\end{figure}

Next consider viscous corrections to HBT radii. The HBT radii are 
calculated with the method of variances. First the ideal
radii parameters  are displayed in Fig. \ref{IdealHBT}. The  results 
are typical of the blast wave parametrization. 
Next the viscous correction to the blast wave results are illustrated in 
Fig. \ref{VisHBTCorrections}.

Viscous corrections to $R_{L}^{2}$ and $R_{O}^{2}$ are large and negative.
This may be understood qualitatively by again considering 
a simple Bjorken expansion of infinite nuclei. The longitudinal 
pressure is reduced, $p_{L} = p - \frac{4}{3}\frac{\eta}{\tau}$. 
Therefore the $p_{z}$ distribution ($
   \frac{dN}{dp_{z} d\eta}$) 
is narrower. However, by boost invariance the single particle distribution ($\frac{dN}{dy\,d\eta}$)  
is a function of $\left|y-\eta\right|$, which yields the relation,
\begin{eqnarray}
   \left. m_{T} \frac{dN}{dp_{z} d\eta}\right|_{\eta=0} 
   = 
   \left.\tau \frac{dN}{dy dz} \right|_{y=0} .
\end{eqnarray}
The $z$ distribution ($\frac{dN}{dy dz} $) at mid rapidity is therefore
narrower because the longitudinal pressure is reduced. This implies
that $R_{L}^2\equiv\langle z^2\rangle$ decreases due to the 
viscous longitudinal
expansion.  In summary, viscosity provides a simple explanation for
the very large radii predicted by ideal hydrodynamics.\\
{\bf Acknowledgments:} This work was supported by DE-AC02-98CH10886.

\end{document}